\documentclass{article}

\usepackage{arxiv}
\usepackage{amsmath}
\usepackage{amsfonts}       % blackboard math symbols
\usepackage{float}
\usepackage{multirow}       % Required for multirow cells in tables
\usepackage[utf8]{inputenc} % allow utf-8 input
\usepackage[T1]{fontenc}    % use 8-bit T1 fonts
\usepackage{hyperref}       % hyperlinks
\usepackage{url}            % simple URL typesetting
\usepackage{booktabs}       % professional-quality tables
\usepackage{amsfonts}       % blackboard math symbols
\usepackage{nicefrac}       % compact symbols for 1/2, etc.
\usepackage{microtype}      % microtypography
\usepackage{lipsum}
\usepackage{graphicx}
\graphicspath{ {./images/} }

\newcommand{\repo}{\url{https://github.com/tlam25/network-of-awards-and-winners}}

\title{Knowledge Graph Enrichment and Reasoning for
Nobel Laureates}

\author{
  Thanh-Lam T. Nguyen \\
  VNU - University of Engineering and Technology \\
  Xuan Thuy, Cau Giay, Hanoi 10000 \\
  \texttt{22024516@vnu.edu.vn} \\
  \And
  Ngoc-Quang Le \\
  VNU - University of Engineering and Technology \\
  Xuan Thuy, Cau Giay, Hanoi 10000 \\
  \texttt{22024510@vnu.edu.vn} \\
  \And
  Thu-Trang Pham \\
  VNU - University of Engineering and Technology \\
  Xuan Thuy, Cau Giay, Hanoi 10000 \\
  \texttt{22024548@vnu.edu.vn} \\
  \And
  Mai-Vu Tran \\
  VNU - University of Engineering and Technology \\
  Xuan Thuy, Cau Giay, Hanoi 10000 \\
  \texttt{vutm@vnu.edu.vn} \\
}

\begin{document}
\maketitle
% Lam
\begin{abstract}
% Mục tiêu dự án
% Phương pháp chính: data augmentation, NER/RE, graph integration, analysis algorithms, chatbot, RAG/GraphRAG
% Kết quả nổi bật
% Đóng góp chính
This project aims to construct and analyze a comprehensive knowledge graph of Nobel Prize and Laureates by enriching existing datasets with biographical information extracted from Wikipedia. Our approach integrates multiple advanced techniques, consisting of automatic data augmentation using LLMs for Named Entity Recognition (NER) and Relation Extraction (RE) tasks, and social network analysis to uncover hidden patterns within the scientific community. Furthermore, we also develop a GraphRAG-based chatbot system utilizing a fine-tuned model for Text2Cypher translation, enabling natural language querying over the knowledge graph. Experimental results demonstrate that the enriched graph possesses small-world network properties, identifying key influential figures and central organizations. The chatbot system achieves a competitive accuracy on a custom multiple-choice evaluation dataset, proving the effectiveness of combining LLMs with structured knowledge bases for complex reasoning tasks. Data and source code are available at: \repo{}.
\end{abstract}

% keywords can be removed

% Trang
\section{Introduction}
% Vấn đề nghiên cứu: xây dựng knowledge graph về Nobel, phân tích mạng, và chatbot graph-aware
% Tại sao bài toán quan trọng
% Tóm tắt ngắn các đóng góp:
    % Ví dụ:
    % - Build enriched Nobel-prize knowledge graph using automatic NER/RE.
    % - Conduct comprehensive social network analysis (small-world, centrality, community detection).
    % - Develop GraphRAG-based chatbot with Text2Cypher finetuning.
Knowledge graphs and advanced network analysis have become essential tools for transforming unstructured text into machine-interpretable insights, enabling the discovery of hidden patterns within complex data. In the domain of scientific and cultural history, the Nobel Prize ecosystem represents a century of interconnected milestones; however, much of this information remains fragmented across static sources like Wikipedia. This lack of structure hinders deep relational analysis and limits the potential for intelligent systems to perform multi-hop reasoning over the data.

To address this challenge, we first establish a rigorous framework for constructing an enriched Nobel knowledge graph. By implementing an automated pipeline for NER and RE task, we transform unstructured biographical texts into a high-coverage graph. This process leverages the capabilities of modern Large Language Models to ensure precise schema-guided extraction, effectively capturing the diverse entities—from laureates to their works and affiliations—that define the domain.

Building upon this structured foundation, we conduct a comprehensive social network analysis to characterize the topology of the Nobel ecosystem. Through the application of centrality measures, community detection algorithms, and small-world statistics, we move beyond simple data storage to uncover the dynamics of scientific collaboration. This analysis reveals the network’s underlying community structures and identifies key cross-disciplinary hubs, offering a quantitative perspective on how prestige and knowledge flow within the network.

Finally, to bridge the gap between complex graph data and end-users, we develop a graph-aware chatbot powered by GraphRAG and a fine-tuned text-to-Cypher model. This system synergizes the reasoning capabilities of LLMs with the precise retrieval of graph databases, allowing users to query the Nobel network using natural language. By enabling accurate handling of multi-hop questions, this work demonstrates a unified approach to building, analyzing, and interacting with domain-specific knowledge graphs.

% Quang
\section{Related Work}
% Tìm vài bài báo giai đoạn từ 2024 tới giờ với các chủ đề sau, viết ngắn gọn người ta là gì và có nhược điểm gì mình giải quyết được không --> nhớ cite
    % Knowledge Graph Construction
    % NER/RE from Wikipedia
    % Social Network Analysis trên scientific communities
    % Text2Cypher, KG-QA
    % RAG và GraphRAG
Recent advancements have shifted from supervised pipelines to LLM-based knowledge graph construction. \cite{zhang2024sackg} introduced SAC-KG, a framework that fine-tunes LLMs to act as skilled automatic constructors for domain-specific knowledge graphs, achieving notable improvements over zero-shot prompting. While effective, generative approaches of this kind may still face challenges in consistently aligning model outputs with a predefined ontology, especially when strict schema adherence is required.

In the realm of extraction from open text, \cite{zhou2024universalner} proposed UniversalNER, which leverages targeted distillation from proprietary LLMs into smaller open-source models to achieve state-of-the-art performance in open-domain Named Entity Recognition. While highly effective for broad sources such as Wikipedia, these distilled models may face challenges when applied to scientific literature, where entities are more domain-specific and often exhibit complex, nested structures.

Analyzing scientific collaboration has evolved beyond simple co-authorship statistics. \cite{amanbek2024resilience} applied complex network analysis to evaluate the resilience and clustering coefficients of collaboration networks within young universities. While such studies provide valuable topological insights, they typically emphasize structural metrics (e.g., centrality or connectivity) and pay less attention to the semantic content underlying collaborations. In contrast, our approach integrates topic modeling directly into the graph, enabling an analysis that captures not only \textit{who} collaborates, but also \textit{what} topics shape and motivate these connections.

Bridging natural language and graph databases remains a critical challenge. \cite{rogge2024realtime} developed a modular framework for real-time Text-to-Cypher generation using LLMs, emphasizing syntax correction and prompt engineering for NoSQL graph databases. While effective for handling general query patterns, current Text-to-Cypher models may still struggle with complex, multi-hop schemas, often producing queries that are syntactically correct but semantically misaligned with the underlying graph structure. To address this, we propose a retrieval-augmented schema injection method designed to improve logical consistency and reduce such semantic errors.

The integration of knowledge graphs into Retrieval-Augmented Generation (RAG) has gained significant traction. \cite{edge2024from} introduced GraphRAG, a method that leverages graph community summarization to answer global, query-focused questions that standard RAG pipelines struggle to resolve. While effective for capturing high-level relational structure, this global summarization strategy can be computationally demanding and may require substantial token budgets during inference.

% Trang
\section{Data Construction}
% Mỗi phần này tạo \subsection rồi mô tả
    % Thu thập dữ liệu Wikipedia
    % Làm sạch, rút gọn, chuẩn hóa
    % Gán nhãn tự động bằng Gemini
    % Schema NER/RE
    % Format JSONL

\begin{figure}[H]
    \centering
    \includegraphics[width=\textwidth]{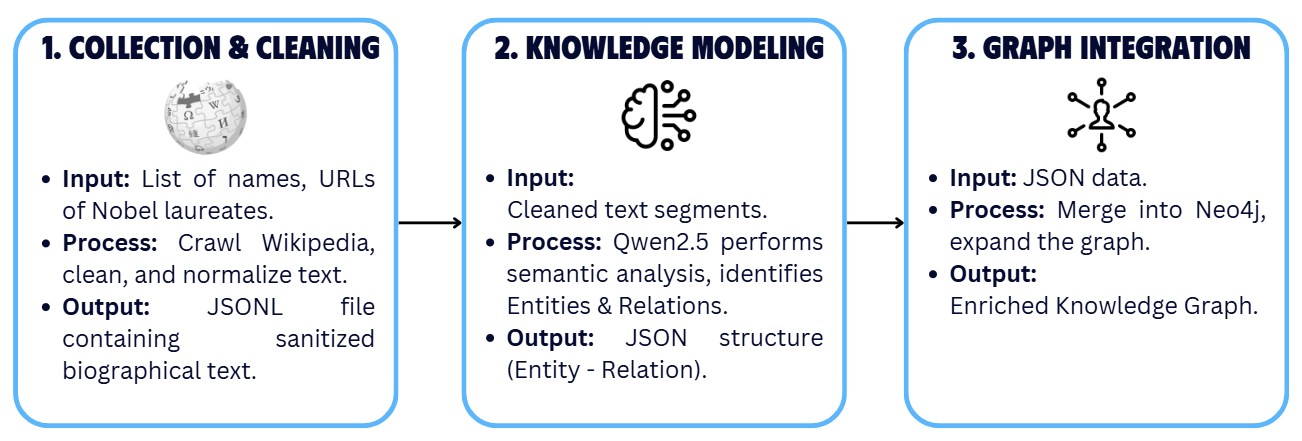}
    \caption{The proposed data construction and enrichment pipeline.}
    \label{fig:pipeline-augment}
\end{figure}

As illustrated in Figure~\ref{fig:pipeline-augment}, the data construction pipeline for building and enriching the Nobel Knowledge Graph operates through four main stages: Wikipedia data collection, text cleaning and normalization, automatic annotation using a large language model, and formatting the final dataset in JSONL for downstream training and graph integration.

\subsection{Data Collection and Cleaning}

We gather biographical text from English Wikipedia for all individuals and organizations that have received Nobel Prizes across six categories: Physics, Chemistry, Physiology or Medicine, Literature, Peace, and Economics. For each target entity, the system identifies the corresponding Wikipedia page via name, QID, or URL, and extracts the lead section and additional high-value narrative content while excluding uninformative parts such as reference lists or navigation templates.

\subsection{Knowledge Modeling}
\paragraph{Text Pre-processing.}
Raw text undergoes a rigorous preprocessing pipeline to eliminate noise and standardize formatting. The process begins with the removal of metadata and non-content navigation elements, followed by the deletion of bracketed artifacts such as citation markers (e.g., \texttt{[1]}, \texttt{[citation needed]}) and the normalization of whitespace and special characters. Subsequently, we employ intelligent sentence segmentation to truncate each entry to a maximum of 10 sentences. Based on our statistics, this number provides a threshold for data richness while maintaining token efficiency for model training. The resulting text segments are clean, coherent, and consistent in length, making them suitable for LLM-based extraction.

\paragraph{Automatic Annotation with Gemini.}
To obtain NER and RE labels at scale without manual annotation, we leverage the Gemini 2.5 Flash API to perform automatic structured extraction. The model is prompted with a predefined schema and produces JSON-formatted outputs containing all detected entities and relations. This approach enables the creation of thousands of high-quality labeled samples, which are subsequently used to fine-tune the Qwen model for domain-specific information extraction.

\paragraph{NER and Relation Extraction Schema.}
We design a comprehensive schema tailored for Nobel-related knowledge, consisting of:

\textbf{Entities:}  
Person, Person\_Non\_Laureate, Organization, Position, Occupation, Field, Country, Location, Award, Notable\_Work, Event.

\textbf{Relations:}  
RECEIVED, WORKS\_AS, WORKS\_IN\_FIELD, EMPLOYED\_BY, EDUCATED\_AT, IS\_CITIZEN\_OF, HOLDS\_POSITION, FOUNDED, CO\_FOUNDED, CO\_DISCOVERED\_WITH, PARTICIPATED\_IN, IS\_SPOUSE\_OF, DEVELOPED.

The schema ensures consistency across annotation and fine-tuning stages and defines the structural backbone of the resulting knowledge graph.

\paragraph{JSONL Formatting}
All processed samples are stored in JSONL format, where each line contains:

\begin{verbatim}
{
  "name": "...",
  "text": "...",
  "entities": [...],
  "relations": [...]
}
\end{verbatim}

This format provides easy integration with LLM training pipelines, compatibility with Neo4j import workflows and flexibility for future dataset expansion.

\subsection{Graph Integration}

The graph integration pipeline consolidates the extracted structured data into the Neo4j database to establish a comprehensive and consistent knowledge network. To preserve the integrity of the existing data, we employ an additive integration strategy that enforces unique constraints on canonical identifiers to prevent duplication. During the ingestion process, the system verifies the prior existence of each entity; existing nodes are updated with new attributes and missing relationships, while new nodes are instantiated for previously unrecorded concepts, ensuring that the original graph structure remains intact.

This integration significantly expands the schema by incorporating novel entity classes such as \texttt{Notable\_Work}, \texttt{Event}, \texttt{Location}, and introducing diverse relationship types including \texttt{CO\_DISCOVERED\_WITH}, \texttt{FOUNDED}, and \texttt{IS\_SPOUSE\_OF}. Ultimately, this transformation evolves the dataset from a simple award registry into a multidimensional social–academic network optimized for complex reasoning and historical analysis.

\subsection{Evaluation Dataset Construction}
To benchmark the system, we generated a comprehensive multiple-choice dataset comprising over $2,000$ questions. This dataset was constructed by sampling sub-graphs and using Gemini 2.5 Flash to paraphrase questions and generate distractors.

\begin{figure}[H]
    \centering
    \includegraphics[width=0.4\textwidth]{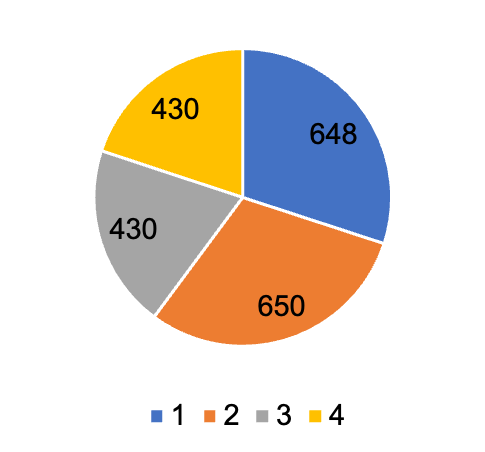}
    \caption{Distribution of Multiple-Choice Evaluation Dataset by Hops}
    \label{fig:mcp_stats}
\end{figure}

The complexity distribution of this evaluation set is depicted in Figure \ref{fig:mcp_stats}. The dataset maintains a balanced distribution across different difficulty levels. Notably, the 4-hop category contains a substantial number of samples ($650$ questions), which is crucial for rigorously testing the system's ability to handle deep reasoning tasks—an area where traditional RAG systems often fail.

\section{Methodology}
% Phần này là phần chính giống report docs của mình

\subsection{Knowledge Graph Enrichment Pipeline}
% Trang
    % Integration Strategy
    % Schema Expansion
    % Duplicate Filtering
    % Final Graph Statistics

The enrichment pipeline robustly integrates automatically extracted entities and relations into the existing Nobel Knowledge Graph while preserving consistency and preventing duplication. The process relies on three main components: a controlled integration strategy, schema expansion to accommodate newly discovered concepts, and duplicate filtering to maintain graph cleanliness. Collectively, these components transform the original Nobel graph into a substantially richer and more expressive knowledge network.

\paragraph{Integration Strategy.}

To ensure stable and conflict-free data ingestion, we define unique constraints on key identity attributes, including names and canonical identifiers (e.g., QIDs) in Neo4j. During the ingestion of JSONL-based NER/RE outputs, the system checks each entity for prior existence in the graph. If an entity exists, we update its attributes and append only missing relationships; if not, we create a new node with the appropriate label. This additive strategy guarantees that no original nodes or edges are overwritten or deleted, preserving the integrity of the initial knowledge base.

\paragraph{Schema Expansion.}

The extraction process yields numerous entity and relation types absent from the original KG schema. To represent this newly surfaced knowledge, we extend the schema with additional entity classes such as \texttt{Person\_Non\_Laureate}, \texttt{Notable\_Work}, \texttt{Event}, and \texttt{Location}. Concurrently, we introduce new relationship types including \texttt{IS\_SPOUSE\_OF}, \texttt{PARTICIPATED\_IN}, \texttt{FOUNDED}, \texttt{CO\_FOUNDED}, \texttt{CO\_DISCOVERED\_WITH}, and \texttt{DEVELOPED}. These extensions broaden the representational capacity of the graph, transforming it from a prize-centered dataset into a multidimensional social--academic network capable of capturing collaborations, career trajectories, institutional affiliations, historical events, and scientific outputs.

\paragraph{Duplicate Filtering.}

Text-based extraction frequently introduces name variants or fragmented entity mentions. To address this, the pipeline employs canonical name normalization to remove parenthetical descriptors and harmonize surface forms. We match entities using rule-based strategies that combine name and entity type to avoid ambiguous merges. Before inserting any relationship, the system verifies that an identical edge (same node pair and relation type) does not already exist. This multi-stage filtering process ensures a clean, non-redundant, and internally consistent graph.

\paragraph{Final Graph Statistics.}

Post-enrichment, the Nobel knowledge graph exhibits a marked increase in both scale and semantic density. As detailed in Table~\ref{tab:graph_stats}, the final graph consolidates $13,666$ nodes and $31,06$3 edges, representing a substantial augmentation of the initial dataset. The network now spans 12 node types, merging core entities (e.g., \texttt{Person}, \texttt{Award}, \texttt{Organization}) with newly discovered classes such as \texttt{Notable\_Work}, \texttt{Event}, and \texttt{Location}. Furthermore, the inclusion of 15 distinct relationship types---including complex interactions like \texttt{PARTICIPATED\_IN}, \texttt{CO\_DISCOVERED\_WITH}, and \texttt{DEVELOPED}---results in a highly connected structure optimized for multi-hop reasoning and graph-aware analysis.

\begin{table}[h]
\centering
\caption{Impact of the enrichment pipeline on graph statistics.}
\label{tab:graph_stats}
\begin{tabular}{lrrrr} 
\toprule
\textbf{Metric} & \textbf{Original} & \textbf{Added} & \textbf{Final} & \textbf{Growth (\%)} \\ 
\midrule
\textbf{Nodes} & 5,437  & 8,229  & \textbf{13,666} & +151.4\% \\
\textbf{Edges} & 19,879 & 11,184 & \textbf{31,063} & +56.3\% \\ 
\bottomrule
\end{tabular}
\end{table}

\subsection{Social Network Analysis Pipeline}
% Lam
    % Graph construction (nodes, edges, weights)
    % Small-world analysis
    % Centrality (PageRank, Degree centrality, Betweenness centrality)
    % Community detection (Louvain)

To analyze the structure of the Nobel community, we constructed an undirected weighted graph $G = (V, E)$ where nodes $V$ represent Laureates and Organizations. The weight $w_{ij}$ of an edge between two nodes $i$ and $j$ is calculated based on the sum of shared attributes (Organization, Field, Country, Award Statement) as follows:
\begin{equation}
    w_{ij} = \sum_{k \in \mathcal{A}} \mathbb{I}(a_i^k = a_j^k)
\end{equation}
where $\mathcal{A}$ is the set of attributes and $\mathbb{I}(\cdot)$ is the indicator function.

\textbf{Small-world Analysis:} To verify the small-world~\cite{watts1998collective} property, we calculated the Average Shortest Path Length ($L$) and the Average Clustering Coefficient ($C$):
\begin{equation}
    L = \frac{1}{N(N-1)} \sum_{i \neq j} d(i, j), \quad C = \frac{1}{N} \sum_{i=1}^{N} \frac{2e_i}{k_i(k_i-1)}
\end{equation}
where $N$ is the number of nodes, $d(i, j)$ is the shortest path distance, $k_i$ is the degree of node $i$, and $e_i$ is the number of links between neighbors of $i$.
\begin{itemize}
    \item The network exhibits $L \approx 1.5$ and $C \approx 0.86$.
    \item Compared to a random graph of similar size ($L_{rand} \approx 1.44, C_{rand} \approx 0.56$), our network has a comparable path length but a significantly higher clustering coefficient ($C \gg C_{rand}$). This confirms the "small-world" property, implying that Nobel laureates form tightly knit clusters linked by short paths.
\end{itemize}

\textbf{Centrality Measures:} We employed three key metrics to rank nodes:
\begin{itemize}
    \item \textbf{PageRank~\cite{brin1998anatomy}:} Identifies influential figures based on connection quality, defined recursively as:
    \begin{equation}
        PR(i) = \frac{1-d}{N} + d \sum_{j \in M(i)} \frac{PR(j)}{Deg(j)}
    \end{equation}
    \item \textbf{Degree Centrality:} Measures popularity by counting direct connections: $C_D(i) = \text{deg}(i)$.
    \item \textbf{Betweenness Centrality:} Identifies bridges between communities by quantifying the number of shortest paths passing through a node $v$:
    \begin{equation}
        C_B(v) = \sum_{s \neq v \neq t} \frac{\sigma_{st}(v)}{\sigma_{st}}
    \end{equation}
\end{itemize}

\textbf{Community Detection:} We utilized the Louvain algorithm~\cite{blondel2008fast} to maximize modularity $Q$, identifying clusters of laureates who frequently interact. The modularity is defined as:
\begin{equation}
    Q = \frac{1}{2m} \sum_{i,j} \left[ A_{ij} - \frac{k_i k_j}{2m} \right] \delta(c_i, c_j)
\end{equation}
where $A_{ij}$ is the adjacency matrix weight, $m$ is total edge weight, and $\delta(c_i, c_j)$ indicates if nodes $i, j$ are in the same community.

\subsection{Chatbot System}
% Quang
    % Overall system architecture
    % Mô hình LLM sử dụng
    % Text2Cypher finetuning pipeline
        % Data finetune (Lam)
        % Finetune nnao
    % Multiple choice evaluation dataset xây như thế nào (Lam)

\begin{figure}[H]
    \centering
    \includegraphics[width=\textwidth]{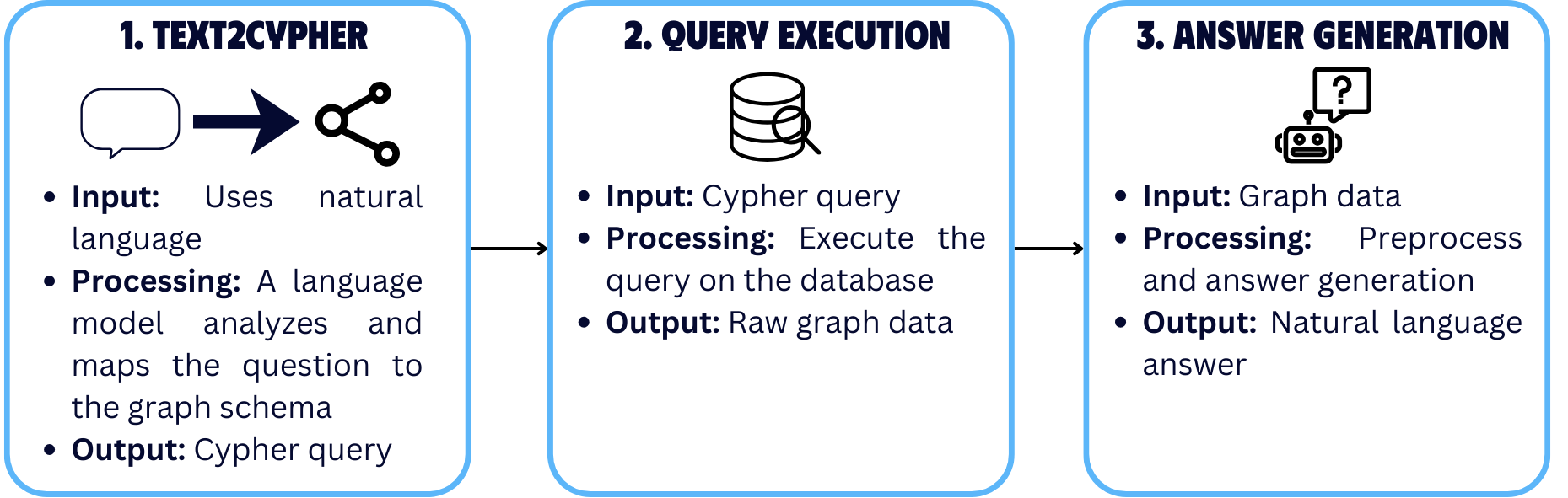}
    \caption{Chatbot Pipeline.}
    \label{fig:chatbot_pipeline}
\end{figure}

As illustrated in Figure~\ref{fig:chatbot_pipeline}, the system consists of three main components. First, the Text2Cypher module converts natural-language questions into Cypher queries aligned with the graph schema. Next, these queries are executed on the graph database and the retrieved results are normalized into contextual input, which the answer generation module then uses to produce accurate and fluent responses.

\subsubsection{Text2Cypher}
The objective of this module is to convert the user’s natural-language question into a Cypher query that is fully aligned with the structure of the underlying knowledge graph. To achieve this, a language model analyzes the input question to identify relevant entities, relation types, and query directions, and subsequently maps these linguistic elements to schema-specific graph patterns. The output of this step is a complete and executable Cypher query, which can be directly processed by the graph database in the subsequent query execution stage. Within the overall pipeline, this module acts as a query translation layer that bridges natural language inputs and graph-structured representations.

\subsubsection{Query Execution}
This component is responsible for executing the Cypher query generated by the previous stage in order to retrieve factual information from the knowledge graph. The system submits the query to the graph database and receives raw results, which may include nodes, relationships, attributes, or lists of values depending on the query semantics. These results faithfully reflect the content of the graph but are not yet optimized for language model consumption. This component serves as the primary data source, providing accurate and structured information for the downstream answer generation stage.

\subsubsection{Answer Generation}
The objective of this stage is to produce a natural, accurate, and user-friendly response based on the information retrieved from the graph database. The raw query results are first cleaned, normalized, and reformatted into a coherent contextual representation. This structured context is then provided to a language model, which interprets the information and generates a fluent answer. The final output may take the form of a complete natural-language response or a structured JSON output for multiple-choice tasks. As the final stage of the pipeline, this component transforms graph-based data into high-quality responses tailored to user needs.

\begin{figure}[H]
    \centering
    \includegraphics[width=0.4\textwidth]{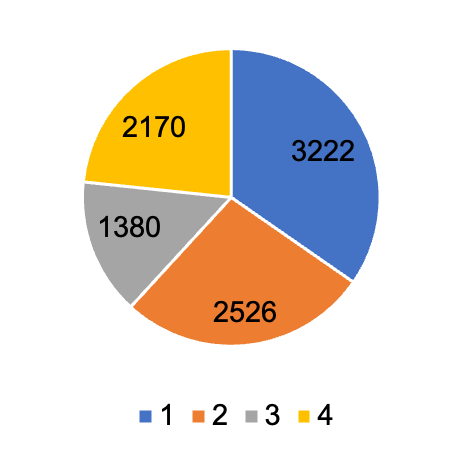}
    \caption{Distribution of Fine-tuning Dataset by Hops}
    \label{fig:finetune_data}
\end{figure}

\section{Experiments and Discussion}

% Trang
\subsection{NER and RE Model Training}
% Loss, token accuracy, entropy…

To enable robust information extraction, we fine-tune the model using a generative text-to-structure paradigm. In this setup, the model receives raw sentences as input and is trained to generate structured JSON outputs representing entity spans and their corresponding relations. This unified formulation allows the model to learn entity boundaries and relational semantics jointly, ensuring stable and schema-compliant extractions suitable for direct integration into the Nobel knowledge graph.

\subsubsection{Fine-tuning Model for NER and RE Task}

\paragraph{The Chosen Model.}
We selected Qwen2.5-0.5B-Instruct as the backbone architecture due to its optimized Transformer design and its extensive context window of 32K tokens, which is particularly well-suited for processing lengthy biographical narratives. A critical advantage of this model is its hardware feasibility; its compact parameter size allows for efficient deployment on mid-range hardware such as the Tesla P100 GPU (16GB VRAM), facilitating rapid experimentation and scalable deployment without the need for the extensive computational resources typically required by larger language models.

\paragraph{Fine-tuning Strategy.}
To achieve efficient adaptation, we employed the QLoRA (Quantized Low-Rank Adaptation)~\cite{dettmers2023qloraefficientfinetuningquantized} technique. Instead of full-parameter fine-tuning, this approach freezes the pre-trained weights and exclusively updates a small set of parameters within low-rank adapter matrices injected into the linear layers of the Attention and MLP blocks. This strategy significantly reduces memory overhead while mitigating the risk of catastrophic forgetting. The training process was conducted over 5 epochs with a learning rate of $3 \times 10^{-4}$, utilizing a Cross-Entropy loss calculated solely on the model's output components to ensure the precise generation of the target JSON structure.

\paragraph{Schema-Constrained Supervision.}
The training data is strictly aligned with the graph schema, incorporating only valid entity types (e.g., \texttt{Person}, \texttt{Organization}, \texttt{Notable\_Work}) and their permissible relations. By exposing the model exclusively to schema-approved labels during the supervised fine-tuning (SFT) phase, we mitigate the risk of hallucinating unsupported entity classes or invalid relationship types. This alignment ensures that the extracted structures are structurally valid and require minimal post-hoc correction before ingestion into Neo4j.

\paragraph{Training Objective.}
The model is optimized using a standard causal language modeling objective. Specifically, we minimize the cross-entropy loss over the target tokens (the JSON output) conditioned on the input instruction and text. This objective encourages the model to internalize the structural constraints of the JSON format while maximizing the likelihood of correctly predicting entity boundaries and relation types defined in the schema. Throughout the training process, we monitor key metrics including training loss, validation loss, and mean token accuracy to ensure the model learns to generate syntactically correct and semantically accurate structures.

\paragraph{Prompt Template.}
Each training instance is formatted as a structured instruction-style prompt comprising the schema definition, the input text, and the target JSON output. The template design forces the model to ground its predictions explicitly in the provided schema, enhancing reliability when processing open-domain Wikipedia text. A representative prompt template is shown below:

\begin{verbatim}
Below is an instruction that describes a task, paired with an input that provides further 
context. 
Write a response that appropriately completes the request.

[INSTRUCTION]
You are an expert in information extraction tasked with building a knowledge graph for 
Nobel Prize laureates. 
I will provide you with the schema of entity types and relation types that need to be 
extracted from the text. Based on the provided schema and text, extract all entities and 
relations associated with the Nobel laureate mentioned in the text and return the output
in JSON format.

[Schema]
{ENTITY TYPES}
{RELATION TYPES}

[TEXT ABOUT {}]
{}

[EXTRACTION RESULTS]
{}
\end{verbatim}

\subsubsection{Training Dynamics and Convergence}

The training trajectory exhibits a stable optimization profile. As shown in our training logs, entropy steadily decreases from $0.57$ to $0.46$, indicating that the model's predictions become increasingly confident and less ambiguous. Concurrently, mean token accuracy improves from $88.18$\% to $88.69$\%, maintaining high consistency throughout the process. Validation loss reaches its minimum at epoch~3 before stabilizing, suggesting that the model has converged to a flat optimal region. The final checkpoint is selected based on a balanced consideration of minimal validation loss and stable token-level accuracy, yielding a reliable extractor for large-scale knowledge graph enrichment.

% Quang
\subsection{Text2Cypher Fine-tuning}

To enable accurate translation from natural language questions to executable Cypher queries, we fine-tune the language model using a text-to-text formulation. In this setting, the model receives a natural-language question together with the full schema of the knowledge graph, and is trained to generate the corresponding Cypher query as output. Providing the complete schema explicitly allows the model to ground its generation process in the actual graph structure, reducing schema violations and improving query correctness.

\subsubsection{Data Generation}
To train the Text2Cypher model, we synthesized a dataset of question-Cypher pairs using varying hop counts ($1$-hop to $4$-hops). We used QLoRA to fine-tune the Qwen model, optimizing for Mean Token Accuracy.

The distribution of the training data is shown in Figure \ref{fig:finetune_data}. As observed, we strictly controlled the data distribution to prevent the model from biasing towards simple queries. Specifically, the dataset is composed of approximately $60$\% $1$-$2$ hop queries and $40$\% $3$-$4$ hop queries. This significant proportion of multi-hop examples (over $3,000$ samples for $3$-$4$ hops) ensures that the model is exposed to sufficient complex reasoning patterns during the training phase.

\subsubsection{Fine-tuning Strategy}

\paragraph{The Chosen Model.}
Qwen3~\cite{qwen3technicalreport} represents the latest generation of large language models in the Qwen series, encompassing both dense and mixture-of-experts architectures. Trained on large-scale and diverse corpora, Qwen3 demonstrates substantial improvements in reasoning ability, instruction following, agent-oriented behavior, and multilingual understanding. A distinctive characteristic of Qwen3 is its unified support for both \emph{thinking} and \emph{non-thinking} modes within a single model, allowing it to dynamically balance complex reasoning tasks and efficient general-purpose interactions.

Compared to earlier models such as QwQ and Qwen2.5-Instruct, Qwen3 exhibits notably enhanced performance in mathematical reasoning, code generation, and commonsense logical inference. In addition, the model shows strong alignment with human preferences, enabling more natural and coherent responses in creative writing, role-playing, and multi-turn conversational settings. Qwen3 also demonstrates advanced agent capabilities, allowing effective integration with external tools and achieving competitive performance on complex agent-based benchmarks among open-source models. Furthermore, the model supports over $100$ languages and dialects, providing robust multilingual instruction-following and translation capabilities.

In this work, we adopt Qwen3-0.6B, a lightweight variant well-suited for efficient deployment while retaining strong reasoning capacity. Its key characteristics are summarized as follows:

\begin{itemize}
\item \textbf{Model Type:} Causal Language Model
\item \textbf{Training Stages:} Pretraining and post-training
\item \textbf{Total Parameters:} $0.6$ billion
\item \textbf{Non-Embedding Parameters:} $0.44$ billion
\item \textbf{Number of Layers:} $28$
\item \textbf{Attention Heads (GQA):} $16$ query heads and $8$ key-value heads
\item \textbf{Maximum Context Length:} $32,768$ tokens
\end{itemize}

This configuration offers a favorable trade-off between computational efficiency and representational capacity, making it particularly suitable for downstream tasks such as Text2Cypher generation and graph-based question answering.

\paragraph{Schema-aware Input Representation.}
Each training instance concatenates the graph schema with the user question in a structured prompt. The schema describes node types, relationship types, and key properties, ensuring that the model has explicit access to all valid graph patterns during generation. This design encourages the model to select only schema-compliant entities and relations when constructing Cypher queries.

\paragraph{Training Objective.}
The fine-tuning process optimizes the standard causal language modeling objective, where the target sequence is the ground-truth Cypher query. Given the schema-augmented prompt, the model learns to map natural-language semantics to graph query patterns, including relation direction, filtering conditions, and aggregation operations.

\paragraph{Prompt Template.}
The prompt used for fine-tuning follows a consistent instruction-based format, as illustrated below:

\begin{verbatim}
You are an expert system that converts natural language questions into Cypher queries.
Use only the schema provided below.
Do not invent new node types or relationships.

[Graph Schema]
{SCHEMA}

[Question]
{USER_QUESTION}

[Cypher Query]
\end{verbatim}

\textbf{Example.}
An example training instance is shown as follows:

\begin{verbatim}
[Graph Schema]
(:Person)-[:IS_CITIZEN_OF]->(:Country)
(:Person)-[:WON_AWARD]->(:Award)

[Question]
Which country has a citizen who won the Nobel Prize in Physics?

[Cypher Query]
MATCH (p:Person)-[:WON_AWARD]->(a:Award {name: "Nobel Prize in Physics"})
MATCH (p)-[:IS_CITIZEN_OF]->(c:Country)
RETURN DISTINCT c.name
\end{verbatim}

This schema-aware fine-tuning strategy enables the model to generate executable and semantically valid Cypher queries, forming a reliable foundation for downstream graph-based question answering.

% Lam
\subsection{Network Analysis Results}
% (đưa các bảng và biểu đồ SNA vào)

This section details the findings from our network analysis, providing empirical evidence of the Nobel community's structural characteristics.

\subsubsection{Small-World Phenomenon}
To validate the small-world nature of the Nobel network, we compared its Average Shortest Path ($L$) and Clustering Coefficient ($C$) against a generated random graph of equivalent size ($999$ nodes in the largest connected component).

\begin{figure}[H]
    \centering
    \includegraphics[width=\textwidth]{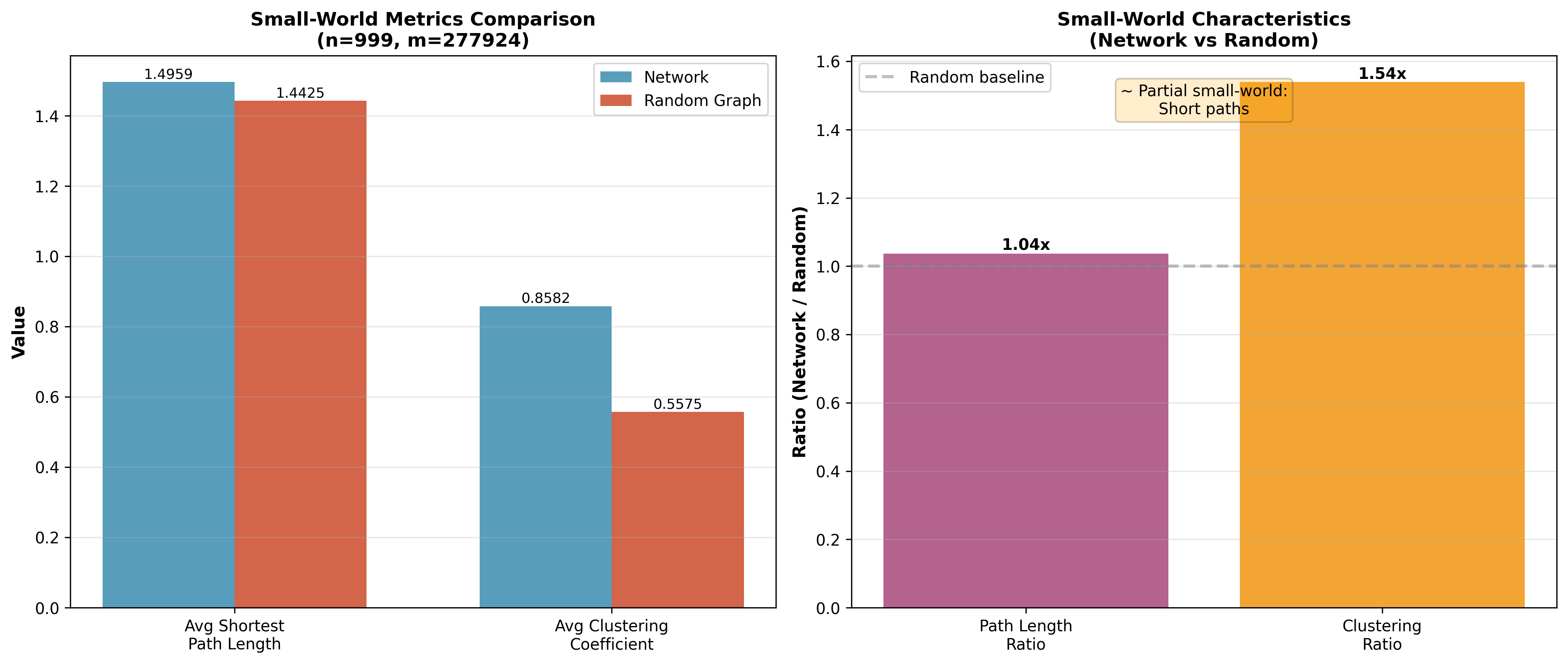}
    \caption{Small-World Metrics Comparison (Network vs Random Baseline)}
    \label{fig:small_world_viz}
\end{figure}

\begin{table}[H]
    \centering
    \caption{Small-World Metric Comparison}
    \begin{tabular}{lccc}
        \toprule
        \textbf{Metric} & \textbf{Target Network} & \textbf{Random Network} & \textbf{Ratio (Target/Random)} \\
        \midrule
        Avg. Shortest Path ($L$) & 1.4959 & 1.4425 & 1.04 \\
        Avg. Clustering Coeff ($C$) & 0.8582 & 0.5575 & 1.54 \\
        \bottomrule
    \end{tabular}
    \label{tab:small_world}
\end{table}

The results in Table \ref{tab:small_world} and Figure \ref{fig:small_world_viz} show that the average path length is very low ($L \approx 1.5$), meaning any two laureates are separated by less than 2 steps on average. This facilitates rapid information flow. Concurrently, the clustering coefficient is significantly higher than random ($0.86$ vs $0.56$), indicating strong local communities (e.g., laureates from the same institution or field tend to know each other). The combination of low $L$ and high $C$ confirms the small-world property.

\subsubsection{Node Ranking and Influence}
We applied three centrality algorithms to identify key figures. The results highlight different types of influence.

\begin{figure}[H]
    \centering
    \includegraphics[width=\textwidth]{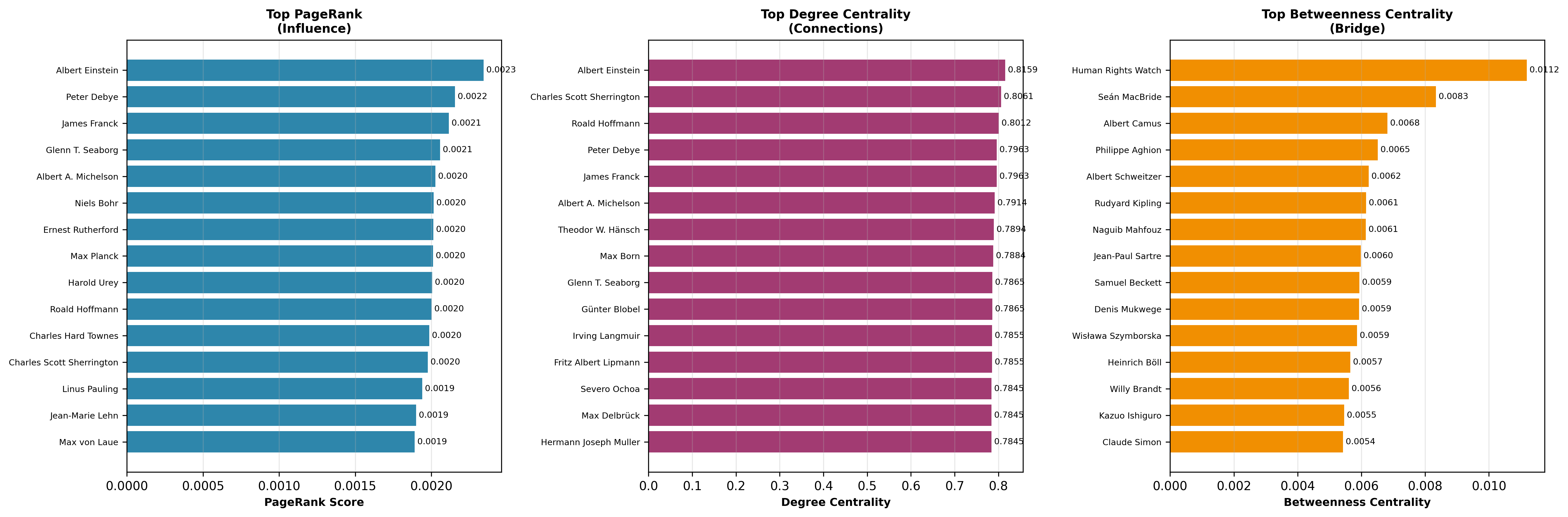}
    \caption{Top Nodes by PageRank, Degree, and Betweenness Centrality}
    \label{fig:ranking_viz}
\end{figure}

\begin{table}[H]
    \centering
    \caption{Top 3 Nodes by Centrality Measures}
    \begin{tabular}{llc}
        \toprule
        \textbf{Algorithm} & \textbf{Name} & \textbf{Score} \\
        \midrule
        \multirow{3}{*}{PageRank (Influence)} & 1. Albert Einstein & 0.0023 \\
        & 2. Peter Debye & 0.0022 \\
        & 3. Glenn T. Seaborg & 0.0021 \\
        \midrule
        \multirow{3}{*}{Degree (Popularity)} & 1. Albert Einstein & 0.8159 \\
        & 2. Charles S. Sherrington & 0.8061 \\
        & 3. Roald Hoffmann & 0.8012 \\
        \midrule
        \multirow{3}{*}{Betweenness (Bridges)} & 1. Human Rights Watch & 0.0112 \\
        & 2. Seán MacBride & 0.0083 \\
        & 3. Albert Camus & 0.0068 \\
        \bottomrule
    \end{tabular}
    \label{tab:centrality}
\end{table}

Albert Einstein dominates both PageRank and degree centrality (see Figure \ref{fig:ranking_viz}). This is not merely due to fame but structural advantage: he had extensive connections (collaborators, shared institutions, field) during a "golden age" of physics, linking him to over $80$\% of the network. Human Rights Watch and Albert Camus top the betweenness centrality. Unlike "hard science" laureates who cluster tightly, Peace organizations and Literature laureates act as bridges. They connect disparate groups (e.g., bridging scientific communities with humanitarian efforts or different nations) that otherwise would not interact directly.

\subsubsection{Community Structure}
Using the Louvain algorithm, we detected $28$ communities with a relatively low modularity of $0.1236$. This low modularity, combined with high graph density, suggests that the Nobel network is a cohesive block with fuzzy boundaries rather than strictly separated groups. We analyzed four distinct community types:

\begin{figure}[H]
    \centering
    \includegraphics[width=\textwidth]{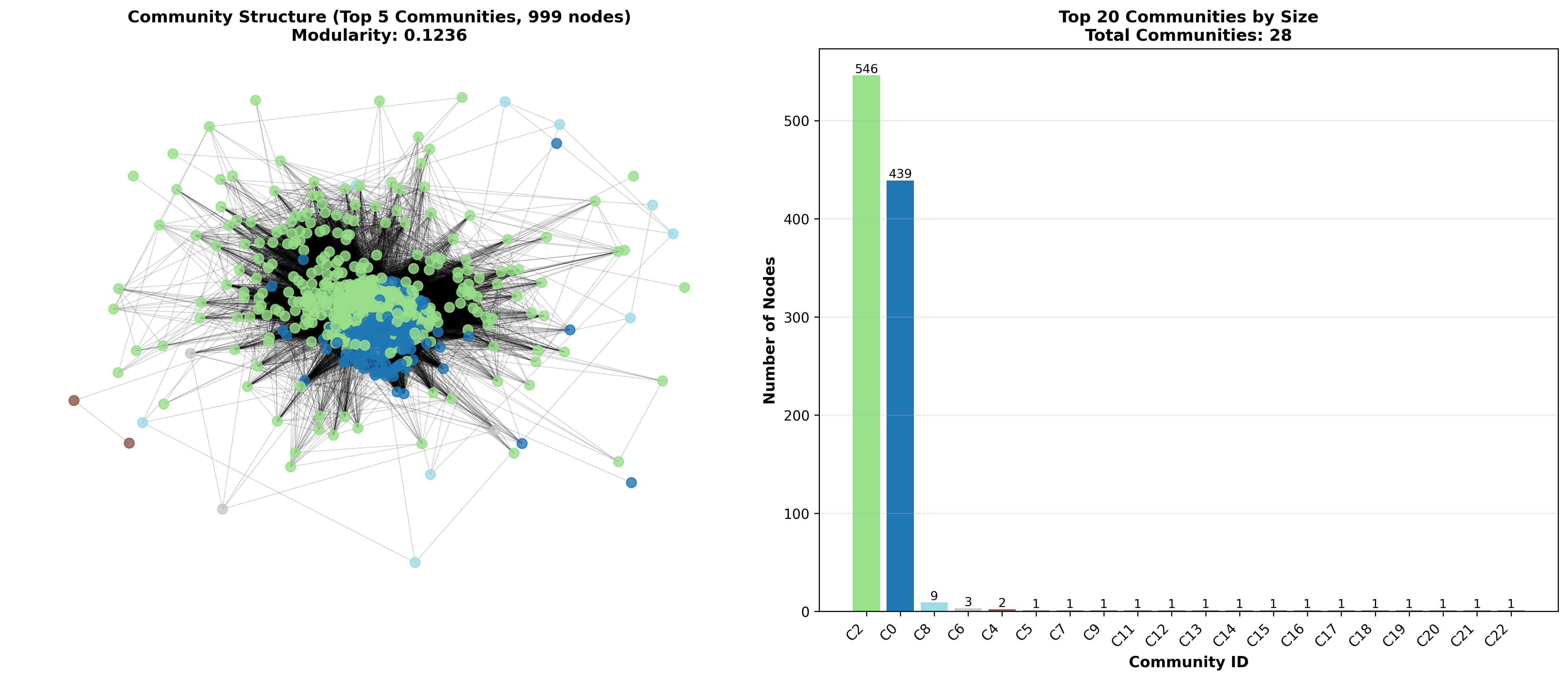}
    \caption{Community Structure Visualization and Size Distribution}
    \label{fig:community_viz}
\end{figure}

\begin{description}
    \item[Community 2 (The "Hard Science" Cluster):] 
    \textit{Size: $546$ nodes.} This is the largest group, containing titans of classical and modern Physics/Chemistry like Marie Curie, Paul Dirac, and Heisenberg. Internal density is moderate ($0.45$), reflecting steady academic collaboration.
    
    \item[Community 0 (The "Social/American" Cluster):]
    \textit{Size: $439$ nodes.} A mix of Economics, Peace, and modern Medicine laureates, including Barack Obama and Martin Luther King Jr. This group has an extremely high internal density ($0.93$), likely driven by the shared attribute "Country: USA", making it a near-complete subgraph.
    
    \item[Community 8 (Humanitarian/NGOs):]
    \textit{Size: $9$ nodes.} A small but crucial cluster containing organizations like the Red Cross and UNHCR. Despite its size, this group has high betweenness, linking international efforts.
    
    \item[Community 6 (Regional/Cultural):]
    \textit{Size: $3$ nodes.} A niche group including Mo Yan, Tu Youyou, and the Dalai Lama. This cluster is formed purely by cultural/geographical ties (China/Tibet), distinct from the Western-centric scientific clusters.
\end{description}

% Quang
\subsection{Chatbot QnA Performance}

Table~\ref{tab:qa_performance} reports the question-answering accuracy of our proposed method compared with Gemini 2.5 Flash Lite, evaluated across different hop levels. Overall, Gemini achieves higher average accuracy ($76.41$\%) than our approach ($72.85$\%), indicating stronger performance on general question answering.

However, a more detailed analysis by reasoning depth reveals distinct behavioral differences between the two systems. Our method demonstrates competitive performance on 2-hop and 4-hop questions, achieving an accuracy of $83.69$\% and $76.98$\%, respectively. In particular, for 4-hop queries, our approach substantially outperforms Gemini ($76.98$\% vs.\ $52.09$\%), suggesting stronger robustness when deeper multi-hop reasoning over the knowledge graph is required.

In contrast, Gemini shows superior performance on 1-hop and 3-hop questions, especially at 3-hop reasoning, where it attains an accuracy of $80.40$\% compared to $58.84$\% for our method. This indicates that large proprietary models may benefit from stronger general reasoning and language priors in certain intermediate-hop scenarios.

Overall, these results suggest that while Gemini excels in shallow and moderately complex reasoning, our graph-based approach maintains more stable performance as the number of reasoning hops increases. This highlights the advantage of explicit knowledge grounding and structured graph traversal for complex multi-hop question answering.

\begin{table}[H]
\centering
\caption{Chatbot QnA performance comparison across different hop levels}
\label{tab:qa_performance}
\begin{tabular}{lcc}
\toprule
\textbf{Metric} & \textbf{Our Method} & \textbf{Gemini 2.5 Flash Lite} \\
\midrule
Accuracy (\%)         & 72.85 & \textbf{76.41} \\
\midrule
Accuracy (1-hop) (\%) & 68.52 & \textbf{79.78} \\
Accuracy (2-hop) (\%) & 83.69 & \textbf{86.46} \\
Accuracy (3-hop) (\%) & 58.84 & \textbf{80.40} \\
Accuracy (4-hop) (\%) & \textbf{76.98} & 52.09 \\
\bottomrule
\end{tabular}
\end{table}

% Lam
\section{Conclusion}
Our project successfully demonstrated an end-to-end pipeline for knowledge graph construction and analysis. By enriching the Nobel Prize dataset with Wikipedia biographies, we uncovered a small-world network structure driven by tight-knit scientific communities and bridged by humanitarian figures. Our GraphRAG chatbot, powered by a fine-tuned Qwen model, offers a cost-effective and accurate solution for querying this complex data, particularly for deep reasoning tasks. Future work will focus on expanding the dataset to include citation networks and further optimizing the Text2Cypher model for even higher accuracy.

\bibliographystyle{unsrt}  
\bibliography{custom}

@inproceedings{zhang2024sackg,
  title={SAC-KG: Exploiting Large Language Models as Skilled Automatic Constructors for Domain Knowledge Graphs},
  author={Zhang, Yijun and others},
  booktitle={Proceedings of the 62nd Annual Meeting of the Association for Computational Linguistics (ACL 2024)},
  year={2024},
  publisher={Association for Computational Linguistics}
}

@inproceedings{zhou2024universalner,
  title={UniversalNER: Targeted Distillation from Large Language Models for Open Named Entity Recognition},
  author={Zhou, Wenxuan and Zhang, Sheng and Gu, Yu and Chen, Muhao and Poon, Hoifung},
  booktitle={The Twelfth International Conference on Learning Representations (ICLR)},
  year={2024}
}

@article{amanbek2024resilience,
  title={Resilience of Scientific Collaboration Networks in Young Universities Based on Bibliometric and Network Analysis},
  author={Amanbek, Yerasyl and Niyazbekova, Shakizada and Nurpeisova, Aigul},
  journal={Data},
  volume={9},
  number={11},
  pages={184},
  year={2024},
  publisher={MDPI}
}

@article{rogge2024realtime,
  title={Real-Time Text-to-Cypher Query Generation with Large Language Models for Graph Databases},
  author={Rogge, Tobias and Weigelt, Sebastian and Tichy, Walter F.},
  journal={Future Internet},
  volume={16},
  number={12},
  pages={438},
  year={2024},
  publisher={MDPI}
}

@article{edge2024from,
  title={From Local to Global: A Graph RAG Approach to Query-Focused Summarization},
  author={Edge, Darren and Trinh, Ha and Cheng, Newman and Bradley, Joshua and Chao, Alex and Mody, Apurva and Truitt, Steven and Larson, Jonathan},
  journal={arXiv preprint arXiv:2404.16130},
  year={2024}
}

@misc{qwen3technicalreport,
      title={Qwen3 Technical Report}, 
      author={Qwen Team},
      year={2025},
      eprint={2505.09388},
      archivePrefix={arXiv},
      primaryClass={cs.CL},
      url={https://arxiv.org/abs/2505.09388}, 
}

@article{watts1998collective,
  title={Collective dynamics of small-world networks},
  author={Watts, Duncan J. and Strogatz, Steven H.},
  journal={Nature},
  volume={393},
  pages={440--442},
  year={1998}
}

@inproceedings{brin1998anatomy,
  title={The anatomy of a large-scale hypertextual Web search engine},
  author={Brin, Sergey and Page, Lawrence},
  booktitle={Proceedings of the 7th World Wide Web Conference (WWW7)},
  year={1998}
}

@article{blondel2008fast,
  title={Fast unfolding of communities in large networks},
  author={Blondel, Vincent D. and Guillaume, Jean-Loup and Lambiotte, Renaud and Lefebvre, Etienne},
  journal={Journal of Statistical Mechanics: Theory and Experiment},
  volume={2008},
  number={10},
  pages={P10008},
  year={2008},
  publisher={IOP Publishing},
  doi={10.1088/1742-5468/2008/10/P10008}
}

@misc{dettmers2023qloraefficientfinetuningquantized,
      title={QLoRA: Efficient Finetuning of Quantized LLMs}, 
      author={Tim Dettmers and Artidoro Pagnoni and Ari Holtzman and Luke Zettlemoyer},
      year={2023},
      eprint={2305.14314},
      archivePrefix={arXiv},
      primaryClass={cs.LG}
}

\end{document}